\documentstyle[12pt]{article}

\author{N. N. Chaus
\\ 
Institute of Mathematics of Ukrainian National Academy \\
of Sciences, vul.Tereshchenkivs'ka, 3, Kiev, 252601, Ukraine,\\
email:chaus@imath. kiev.ua}

\title{On the theory of gravitation field}

\date{ }

\begin{document}

\maketitle

\begin{abstract}
We construct a general relativity formula for the law of gravity for
material bodies. The formula contains three numeric parameters that
are to be determined experimentally. If they  are chosen from
symmetry considerations, then the theory that appears is close to the
theory of electrodynamics: the gravitational field is given by two
vector fields, one can write the energy-momentum tensor, we give an
answer on the question what a gravitational wave is. Going to infinity,
this wave carries with it the negative energy. 
\end{abstract}

\section*{The author's platform}
The author divests himself of the official ideology on the subject
\cite{L.L}.

The vacuum is a physical medium. It can be acted upon, and its
condition can be changed. The gravitational field is a special state,
special deformation of the vacuum, and this field is not the main,
most important physical field. The actual energy, the mass as a
synonym for the energy, does not generate a gravitational field. But it
can be generated by physical bodies, since they have at their
disposal the corresponding mechanism of interaction with the vacuum.
Naturally, elementary particles (not all) should also have this
mechanism. These particles, from their ``birth'', sit on the vacuum
and are accompanied by gravitational fields. The energy by itself has
nothing of the sort. The electron has a mass, but this does not mean
that it possesses its own gravitational field. The author is afraid
that the mystery of the strange constant ${2.4\cdot 10^{-43}}$ could
turn out to be a miscomprehension. The same can be said about
photons, with the remark that there is no experimental data showing
that these particles possess a gravitational field. The author would
consider it a great miracle, if photons passed by the Sun without a
trajectory curving. There could be a dozen of reasons to prohibit
photons to move in a straight line.

The author claims that the gravitational field of a physical body is
determined, first of all, by the quantity and quality of elementary
particles that make the body. This is precisely at this point where
the notion of a heavy mass appears as an integral characteristic of
the body. A commonly accepted in physics hypothesis is that a heavy
and inertial masses are equivalent. This hypothesis does not
contradict to known at this point experimental data. It also looks very
attractive. But the author does not believe that the beauty will save
the world.

In this article, we accept that the notions of a heavy and inertial
mass are unequal. The formulas and conclusions proposed in this
article do not either contradict to any known experimental facts.

We take the following well known formulation as a fundamental one.

Let $\Omega$ be some bounded region in ${R^3}$, and ${\sigma_0}$ be
the density of a distributed in $\Omega$ heavy mass. Let $\varphi$ be
a solution of the equation ${\Delta\varphi=4\pi\kappa
{\sigma_0}}$, and $\varphi\to 0$, $|r|\to\infty$ . Then there exists
$\kappa$ such that the formula ${\bf g}=-{\sigma_0\nabla\varphi}$
gives the force density with which the gravitational field acts on a
heavy mass in an elementary volume. This parameter, $\kappa$, is
called gravitational constant, and the function $\varphi$ is the
potential of the gravitational field.

\section*{Relativistic formulas}

We need to obtain a relativistic formula for the force density ${\bf
g}$ for a moving heavy mass. At this point we suppose that the motion
of the heavy mass can be determined by the action of external,
nongravitational forces. In other words, we suppose that the velocity
field of the heavy mass, ${v_\alpha}$, is a quantity by itself in the
theory.

Denote by $V_k$ the $4$-velocity%
\footnote{
Here and in the sequel, $ V_\alpha=\gamma v_\alpha$, $V_4=ic\gamma$,
$(x_1,x_2,x_3,x_4)=(x,y,z,ict)$, $\partial_n=\partial/\partial x_n$.}
of the particles, and by $\sigma_0$ --- the density of the heavy mass
relatively to an IRF in which the particles are stationary. Let us
form a $4$-vector, ${s_k=\sigma_0V_k}$, and call it the $4$-flux of
the heavy mass. At this point it is almost necessary to replace the
equation ${\Delta\varphi=4\pi\kappa{\sigma_0}}$ with the system
\begin{equation}
\partial_n^2 \Phi_k =-4\pi c^{-2}\kappa s_k,\qquad k=1,2,3,4, \label{1}
\end{equation}

The solution of this system subject to ${\Phi_k(r,t)\to 0,|r|\to \infty,
k=1,2,3,4}$, will be called $4$-potential of the gravitational field
induced by the $4$-flux of the heavy mass $s_k$. In the case where
the masses are stationary, we get ${\partial_k^2
\Phi_4=\Delta\Phi_4=} {-4i\pi c^{-2}\kappa\sigma_0}$,
$\Phi_\alpha\equiv 0$, whence ${\Phi_4=-ic^{-1}\varphi}$,
${g_\alpha=-ic\sigma_0\partial_\alpha\Phi_4}$. It is naturally to
assume that the force we are searching for, ${\bf g}$ (and the
$4$-force $g_k$) should depend, in the general case, only on first
order derivatives $\partial_j\Phi_n$, that it should be proportional
the density $\sigma_0$, depend in a certain way on $V_j$ and, maybe,
on $\partial_nV_j$. To find such a dependence, we use the necessary
condition ${g_kV_k=0}$, which yeilds ${g_k=\Theta_{ks}V_s}$ with some
antisymmetric matrix $\Theta_{ks}$. It is clear that this matrix is
generated by the quantities in the theory and must have a correct
dimension. This leads to the following possible ways to construct the
matrix $\Theta_{ks}$:
\begin{eqnarray*}
\Theta_{ks}(1)&=&-\sigma_0(\partial_k\Phi_s-\partial_s\Phi_k),\\
\Theta_{ks}(2)&=&\sigma_0(p_{sn}\partial_n\Phi_k-p_{kn}
\partial_n\Phi_s),\\
\Theta_{ks}(3)& =&\sigma_0(q_{sn}\partial_k\Phi_n-q_{kn}\partial_s
\Phi_n),\\
\Theta_{ks}(4)&=&\sigma_0(M_{sk}-M_{ks})\partial_n\Phi_n,\\
\Theta_{ks}(5)&=&\sigma_0(N_{skmn}-N_{ksmn})\partial_n\Phi_m.
\end{eqnarray*}

All quantities $p_{sn}$, $q_{sn}$, $M_{sk}$, $N_{skmn}$ in these
formulas are dimensionless. We have no choice but to construct these
expressions by using the dimensionless ${c^{-1}V_{j}}$. Thus we get
$p_{sm}\sim q_{sm}\sim c^{-2}V_sV_m$, $M_{sk}=M_{ks}$,
$N_{skmn}=N_{ksmn}$, so that $\Theta_{ks}(4)=\Theta_{ks}(5)=0$. Thus
three first constructions of the matrix $\Theta_{ks}$ remain, and, up
to a constant factor,
\begin{eqnarray*}
\Theta_{ks}(2)&=&\sigma_0c^{-2}V_n(V_s\partial_n\Phi_k-V_k
\partial_n\Phi_s),\\
\Theta_{ks}(3)&=&\sigma_0c^{-2}V_n(V_s\partial_k\Phi_n-V_k
\partial_s\Phi_n).
\end{eqnarray*}

These three found tensors should be supplemented with the tensor\\
$\Theta_{ks}(1^*)=e_{ksnm}\Theta_{nm}(1)$, where $e_{ksnm}$ is the
totally antisymmetric unit tensor. There is no sense in introducing
similar $\Theta_{ks}(2^*)$ and $\Theta_{ks}(3^*)$, since
$\Theta_{ks}(2^*)V_{s}=0$ and $\Theta_{ks}(3^*)V_{s}=0$. Finally,
there are $4$ choices for the formula of the $4$-density $g_{k}$:
\begin{equation}
\begin{array}{rcl}
g_{k}(1)&=&-\sigma_0V_{s}(\partial_k\Phi_{s}-\partial_s\Phi_{k}),\\[2mm]
g_{k}(2)&=&-\sigma_0V_{n}\partial_n\Phi_k-\sigma_0c^{-2}
V_{n}V_{s}V_{k}\partial_n\Phi_s,\\[2mm]
g_{k}(3)&=&-\sigma_0V_{n}\partial_k\Phi_n-\sigma_0c^{-2}
V_{n}V_{s}V_{k}\partial_s\Phi_n,\\[2mm]
g_k(1^*)&=&-\sigma_0e_{ksnm}V_{s}(\partial_n\Phi_m-\partial_m
\Phi_n).
\end{array}
\label{2}
\end{equation}

\subsection*{Remark}
We could have assumed that the formula for $g_k$ also contains linear
expressions of the potentials $\Phi_n$. Then there would be two more
possible constructions for $\Theta_{ks}$,
$$
\Theta_{ks}(6)=\sigma_0(a_k\Phi_s-a_s\Phi_k),\qquad \Theta_{ks
}(7)=\sigma_0(b_{ksn}-b_{skn})\Phi_n.
$$
Here the quantities $a_k$ and $b_{ksn}$ must have the same dimension
as $c^{-1}\partial_nV_j$. This leads to the formulas (up to a
constant factor):
\begin{eqnarray*}
\Theta_{ks}(6)&=&c^{-2}\sigma_0V_n(\Phi_s\partial_nV_k-\Phi_k%
\partial_nV_s),\\
\Theta_{ks}(7a)&=&c^{-2}\sigma_0V_n\Phi_n(\partial_kV_s%
-\partial_sV_k),\\
\Theta_{ks}(7b)&=&c^{-2}\sigma_0\Phi_n(V_k\partial_nV_s-V_s%
\partial_nV_k),\\
\Theta_{ks}(7c)&=&c^{-2}\sigma_0\Phi_n(V_k\partial_sV_n-V_s%
\partial_kV_n).
\end{eqnarray*}
One should also add $\Theta_{ks}(6^*)=e_{ksnm}
\Theta_{nm}(6)$ to these tensors, since\\
$\Theta_{ks}(6^*)V_s\not\equiv 0$. This collections of tensors gives
the following possibilities for formulas for $g_k=\Theta_{ks}V_s$:
\begin{eqnarray*}
g_k(6)&=&c^{-2}\sigma_0\Phi_sV_sV_n\partial_nV_k,\qquad
g_k(7b)=\sigma_0\Phi_n\partial_nV_k,\\
g_k(7c)&=&\sigma_0\Phi_n(\partial_kV_n+c^{-2}V_kV_s
\partial_sV_n),\qquad g_k(6^*)=\Theta_{ks}(6^*)V_s.
\end{eqnarray*}
We see that all these $g_k$ can not express the force of the
gravitational field, since they all vanish for $v_\alpha =0$. But
these formulas could give an additional term in the formulas
containing $\partial_j\Phi_n$. We do not discuss this subject any
further.

Let us go back to formulas (2). For a heavy stationary mass, we have
\begin{eqnarray*}
g_k(1)=-ic\sigma_0\partial_k\Phi_4, \qquad g_k(2)=0, \qquad
g_k(3)=g_k(1), \qquad g_k(1^*)=0.
\end{eqnarray*}

Thus formulas for $g_k(2)$ and $g_k(1^*)$ could also give additional
terms in formulas for $g_k(1)$ and $ g_k(3)$, and the final formula
for $g_k$ has the form:
\begin{equation} \label{3}
g_k=\lambda g_k(1)+(1-\lambda)g_k(3)+\mu g_k(2)+ \nu g_k(1^*).
\end{equation}
The numbers $\lambda$, $\mu$, and $\nu$ must be determined from
experiments. It is feasible that $\lambda=1$, ${\mu =\nu =0}$. In
such a case, we get the formula:
\begin{equation} \label{4}
g_k=-s_m(\partial_k\Phi_m-\partial_m\Phi_k).
\end{equation}

In the sequel we assume formula (4) to hold.

\subsection*{The fields ${\bf F}$ and ${\bf G}$ of the
gravitational field.}

Formulas (1) and (4) have the form of the main formulas in classical
electrodynamics \cite{A}. And by similarity to electrodynamics,
introduce the vector fields ${\bf F}$ and ${\bf G}$ which, together
with the $4$-potential $\Phi_k$, are characteristics of the
gravitational field. Introduce the notations
${\sigma=\sigma_0\gamma}$, ${{\bf s}=(s_1,s_2,s_3)}$, ${{\bf
\Phi}=(\Phi_1,\Phi_2,\Phi_3)}$, ${\varphi=ic\Phi_4}$, ${{\bf
F}=\nabla \varphi-{\bf\dot \Phi}}$, ${{\bf G}={\rm rot\,}{\bf
\Phi}}$, ${\bf g}=(g_1,g_2,g_3)$. Then formulas (4) give the
following:
$$
{\bf g}=-\sigma{\bf F}-[{\bf s},{\bf G}].
$$

It is easy to check that the following relations hold:
\begin{equation} \label{5}
\begin{array}{c}
{\rm div\,}{\bf G}=0,\qquad {\bf\dot G}+{\rm rot\,}{\bf F}=0,%
\qquad {\rm div\,}{\bf F}=4\pi \kappa\sigma-\partial_k\dot\Phi_k,\\
 {\rm rot\,}{\bf G}-c^{-2}{\bf\dot F}=4\pi\kappa c^{-2}%
{\bf s}+\nabla (\partial_k \Phi_k).
\end{array}
\end{equation}

Let ${\partial_ks_k=0}$. Then the relationship between the fields
${\bf G}$, ${\bf F}$, and ${\bf s}$ will be the same as between the
fields ${\bf B}$, ${\bf E}$, and ${\bf j}$ in classical
electrodynamics \cite{A}. Here the condition ${\partial_ks_k=0}$
corresponds to the electrical charge conservation law in
electrodynamics. In our case, the heavy mass is characterized by the
quantity and quality of the particles that make up the body, and if a
change of its volume induces a change of its energy, the heavy
mass of the body does not need to change. Thus, {\em the condition that
${\partial_ks_k=0}$ is absolutely natural for the $4$-flux $s_k$ of
the heavy mass, and we add it to the general definition of the heavy
mass}.

\subsection*{The energy-momentum tensor}

The main equation for the energy-momentum tensor ${\tau_{ks}}$ of the
gravitational field is the equation ${\partial_s \tau_{ks}=g_k}$ with
the right-hand side given by
${g_k=}{-s_m(\partial_k\Phi_m-\partial_m\Phi_k)}$. The author
believes that there are no reasons for defining the notion of the
energy-momentum tensor of the gravitational field, since the very
notion of the $4$-density $g_k$ is not defined. At this point one
could give a simple answer to this question. The formulas given here
repeat main formulas of electrodynamics, and thus to write down the
tensor ${\tau_{ks}}$, we use the Poynting's tensor
${T_{ks}}$\cite{A}. Denoting ${L_{km}=
\partial_m\Phi_k-\partial_k\Phi_m}$, we get:
$$
\tau_{ks}=\frac{c^2}{4\pi\kappa}L_{km}L_{sm}-\frac{c^2}
{16\pi\kappa}L_{nm}^2\delta_{ks}.
$$

Note that this formula can only be written if ${\partial_ks_k=0}$,
whereas formula (4) was obtained without this condition. The tensor
$\tau_{ks}$ defines the density of the gravitational energy in the
space ${W=\tau_{44}}$, flux of the gravitational energy ${\bf
S}=(S_1,S_2,S_3)$, where ${S_{\alpha}=ic\tau_{4\alpha}}$, and the
density of the momentum ${p_{\alpha}=ic^{-1}\tau_{\alpha 4}}$. It is
not difficult to calculate that
$$
W =-\frac{1}{8\pi\kappa}{\bf F}^2- \frac{c^2}{8\pi\kappa}
{\bf G}^2, \qquad {\bf S}=-\frac{c^2}{4\pi\kappa}
[{\bf F},{\bf G}].
$$
Note that the quantity $W$ is negative. This means that the creation
of the gravitational field  leads to taking energy from the vacuum.
And, consequently, unperturbed vacuum has a certain inner reserve of
energy. This is not something extraordinary, since vacuum is a
medium.

The solution of system (5), ${\bf F}=(0,ca_1(x-ct),
ca_2(x-ct))$, ${\bf G}=\\=(0,-a_2(x-ct),a_1(x-ct))$, is an example
of a flat gravitational wave.\\ It has ${W=\frac{-c^2}{4\pi
\kappa}(a_1^2+a_2^2)}$, ${{\bf S}=(Wc,0,0)}$. Going to infinity, the
wave takes with it the negative energy.

Let, for example, an electric charge oscillate along certain axis.
Then there will be electromagnetic radiation that carries with it
positive energy. The source of this energy is in the cause that keeps
the process oscillating. If a heavy mass will oscillate about an axis,
then, instead of electromagnetic, there will be gravitational
radiation. But the system now looses negative energy, i.e. its total
energy increases, but, of course, not every kind of the energy increases. And
it could be that precisely mechanical oscillations recuperate those
material systems that are not able to radiate electromagnetic waves
any more. Let us also note that experiments on detecting
gravitational waves will hardly be successful if they are oriented
with an expectation that they should carry positive energy. Recall
that we set ${\lambda=1}$, ${\mu=\nu=0}$ in formula (3). For another
choice of ${\lambda,\mu,\nu}$, the theory experiences some
difficulties. The formula for ${g_k(2)}$ contains the third power of
velocity, and thus the tensor ${\tau_{ks}(2)}$ can not be written in
terms of  ${\sigma_0}$, ${\Phi_n}$, ${V_n}$ and their derivatives with
the condition that ${\partial_s \tau_{ks}(2)=g_k(2)}$. It is also
impossible to write the tensor ${\tau_{ks}(1^*)}$ subject to
${\partial_s \tau_{ks}(1^*)=g_k(1^*)}$ because of a poor geometry of
the formula for ${g_k(1^*)}$.

\end{document}